\documentclass[reprint,twocolumn,aps,prl,amsmath,amssymb,floatfix,superscriptaddress,longbibliography]{revtex4-2}
\usepackage{graphicx}
\usepackage{epsfig}
\usepackage{bm}
\usepackage{dcolumn}
\usepackage{color}
\usepackage{physics}
\usepackage{float}
\usepackage{makecell} 
\usepackage{booktabs} 
\usepackage{amsmath}
\usepackage[colorlinks,urlcolor=cyan,citecolor=blue,linkcolor=magenta]{hyperref}
\makeatletter
\newcommand*{\rom}[1]{\expandafter\@slowromancap\romannumeral #1@}
\makeatother
\usepackage{tikz}
\usepackage{subfigure}

\begin{document}
	
\preprint{This line only printed with preprint option}

\title{Anderson-skin dualism: A boundary-dependent effect in non-Hermitian disordered coupled systems}

\author{Shan-Zhong Li}
\affiliation {Key Laboratory of Atomic and Subatomic Structure and Quantum Control (Ministry of Education), Guangdong Basic Research Center of Excellence for Structure and Fundamental Interactions of Matter, School of Physics, South China Normal University, Guangzhou 510006, China}
\affiliation {Guangdong Provincial Key Laboratory of Quantum Engineering and Quantum Materials, Guangdong-Hong Kong Joint Laboratory of Quantum Matter, Frontier Research Institute for Physics, South China Normal University, Guangzhou 510006, China}

\author{Linhu Li}
\email[Corresponding author: ]{lilinhu@quantumsc.cn}
\affiliation {Quantum Science Center of Guangdong-Hong Kong-Macao Greater Bay Area, Shenzhen 518048, China}

\author{Shi-Liang Zhu}
\affiliation {Key Laboratory of Atomic and Subatomic Structure and Quantum Control (Ministry of Education), Guangdong Basic Research Center of Excellence for Structure and Fundamental Interactions of Matter, School of Physics, South China Normal University, Guangzhou 510006, China}
\affiliation {Guangdong Provincial Key Laboratory of Quantum Engineering and Quantum Materials, Guangdong-Hong Kong Joint Laboratory of Quantum Matter, Frontier Research Institute for Physics, South China Normal University, Guangzhou 510006, China}
\affiliation{Quantum Science Center of Guangdong-Hong Kong-Macao Greater Bay Area, Shenzhen 518048, China}

\author{Zhi Li}
\email[Corresponding author: ]{lizphys@m.scnu.edu.cn}
\affiliation {Key Laboratory of Atomic and Subatomic Structure and Quantum Control (Ministry of Education), Guangdong Basic Research Center of Excellence for Structure and Fundamental Interactions of Matter, School of Physics, South China Normal University, Guangzhou 510006, China}
\affiliation {Guangdong Provincial Key Laboratory of Quantum Engineering and Quantum Materials, Guangdong-Hong Kong Joint Laboratory of Quantum Matter, Frontier Research Institute for Physics, South China Normal University, Guangzhou 510006, China}

\date{\today}
\begin{abstract}
We report a novel localization phenomenon that emerges in non-Hermitian and quasiperiodic coupled systems, which we dub ``Anderson-Skin (AS) dualism". The emergence of AS dualism is due to the fact that non-Hermitian topological systems provide non-trivial topology for disordered systems, causing the originally localized Anderson modes to transform into skin modes, i.e., the localized states within the point gap regions have dual characteristics of localization under periodic boundary condition (PBC) and skin effects under open boundary conditions (OBC). As an example, we analytically prove the 1D AS dualism through the transfer matrix method. Moreover, by discussing many-body interacting systems, we confirm that AS dualism is universal.
\end{abstract}

\maketitle

\textcolor{blue}{\emph{Introduction.}}---The study of open (non-conservative) systems, as a direction that is highly concerned in multiple fields such as condensed matter physics, complex systems and phase transitions, and atomic and molecular optics, etc., has long been the research focus of physicists~\cite{VVKonotop2016,RElGanainy2018}. The non-Hermitian Hamiltonian, as a powerful tool for describing problems in open systems, has been widely used in recent years. In the non-Hermitian systems described by the non-Hermitian Hamiltonian, in addition to some phenomena existing in Hermitian systems, there will also emerge some phenomena that are completely absent in Hermitian situations, such as Exceptional point/Spawning rings~\cite{WDHeiss2012,MAMiri2019,KDing2022,BZhen2015}, Skin effect~\cite{SYao2018,ZGong2018,KKawabata2019,LLi2020,KZhang2022}, etc. These phenomena have also been successively realized in various experimental platforms such as ultracold atoms~\cite{MNakagawa2018,WGuo2020,QLiang2022,JLi2019,ZRen2022,EZhao2025}, photonic lattices~\cite{LFeng2017,MPan2018,SWeidemann2020,ZFang2022,WWang2023}, and topological quantum circuits~\cite{LSu2023,THelbing2020,XZhang2021,CXGuo2024,LLi2025,DHalder2025,CHLee2018}. The main reason for the emergence of the novel phenomena is that complex spectra exist in non-Hermitian systems, which can also lead to a novel phenomenon that is unique in non-Hermitian systems, i.e., topological point gaps in the complex plane~\cite{ZGong2018,KZhang2020,KKawabata2019,DSBorgnia2020}. The emergence of these point gaps is accompanied by a sudden change in the spectral winding number from $0$ to $1$. Besides, the distribution of eigenvalues under OBC will form a gapless spectrum without loop structure, and the corresponding eigenstates that fall within the point-gap-loop of PBC will be localized at the boundaries, that is to say, the skin effect will emerge~\cite{KZhang2020}. This connection between the point-gap spectrum and the skin effect has been confirmed in 1D~\cite{KZhang2020,SLonghi2019,JClaes2021}, 2D~\cite{KZhang2022,CHLee2019,KKawabata2020,YLi2022}, and many-body systems~\cite{LJZhai2020,HZLi2023,KKawabata2022,TYoshida2024,HJGliozzi2024}.


On the other hand, Anderson localization, as an old yet profoundly influential physics phenomenon, remains an important branch in condensed matter physics and semiconductor physics~\cite{PWAnderson1958,EAbrahams1979,PALee1985,FEvers2008,ALagendijk2009}. Unlike the edge-localized modes produced by non-Hermitian skin effects, Anderson localization is the localization in bulk caused by quantum interference due to disorder. Recently, the competition between Anderson localization and non-Hermitian skin localization has attracted extensive attention~\cite{YLiu2021,HJiang2019,ZGong2018,XCai2021,XCai2022,LZTang2021,LZTang2022,SLJiang2023,ZQZhang2023,HLiu2023,SLonghi2019a,XPJiang2024,LWang2024a,LWang2024b,LJZhai2020,APAcharya2024,TLiu2020,XTong2025,SZLi2024,SZLi2024a,HZLi2025,YLiu2021a}. In non-reciprocal disordered systems, the localization phase transition is accompanied by the closing of point gaps, and the corresponding spectral winding number also changes from non-zero to zero~\cite{HJiang2019}. 

So far, the common understanding is that Anderson localized states can only appear outside the region of the point-gap-loop spectrum and are not affected by boundary conditions, with their corresponding spectral winding numbers always being zero~\cite{HJiang2019,SZLi2024,YLiu2021}. This raises a critical question: Can localized states emerge within the point-gap-loop? If so, then what unique characteristics do these localized states possess? 
 \begin{figure}[tbhp]
	\centering	\includegraphics[width=8cm]{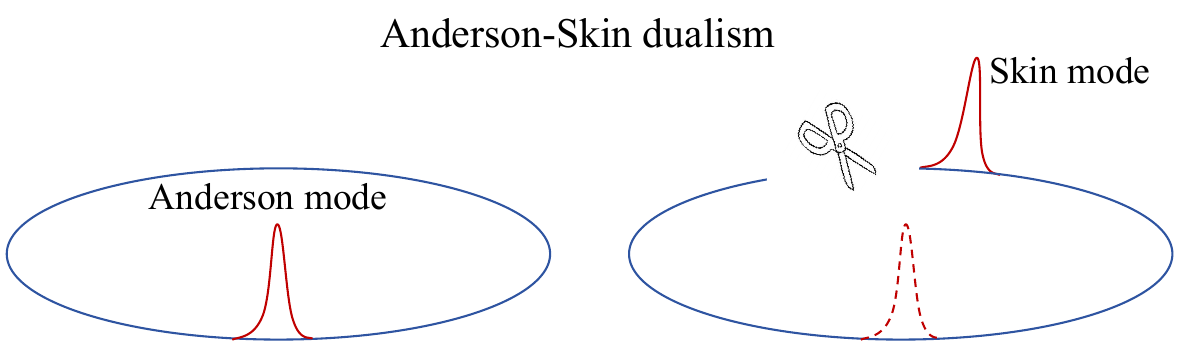}
	\caption{Schematic diagram of AS dualiam.}\label{F1AS}
\end{figure}

In this Letter, we study non-Hermitian and quasiperiodic coupled systems. We demonstrate that the eigenvalues corresponding to localized states can indeed emerge within the region of point-gap-loops. Moreover, these states exhibit a very unique boundary-dependent phenomenon, namely AS dualism [see Fig.~\ref{F1AS}]. We have summarized the corresponding characteristics of AS dualism states in Tab.~\ref{T1ASC}.

\begin{table}[htbp]
\setlength{\tabcolsep}{5pt} 
	\centering
	\caption{The differences among Skin effect, Anderson localizaton, and AS dualism.}
	\begin{tabular}{lccc}

		\toprule

		  Modes       & Skin      & Anderson     & AS dualism                      \\ 
		\midrule\midrule
		Origin   & \makecell{Spectrum\\point gap} & \makecell{Wave\\ interference} & \makecell{Point gap \\ and interference} \\ 
		\midrule
		Position & Boundary  & Bulk              & \makecell{Bulk (PBC)\\ Boundary (OBC)}       \\ 
  	   \midrule
		\textbf{$\omega(E_{b})$} & $\ne 0$  & $0$              & {$\ne 0$}       \\ 
  
		\bottomrule
	\end{tabular}\label{T1ASC}
	\small 
\end{table}

\textcolor{blue}{\emph{AS dualism of non-interacting systems}}---Different from the traditional global topological phase, the non-Hermitian multi-band model can form the point-gap structure for some bands, while the remaining bands are in the gapless phase. Therefore, to achieve AS dualism, we need to realize Anderson localized states surrounded by a spectral topological point gap, thereby achieving the boundary-controlled localization behavior mentioned in Fig.~\ref{F1AS}. A simple example is the two-chain model, where one chain provides non-trivial point gaps while the other chain provides Anderson localized states. As a representative example, we here utilize a 1D coupled system of Hatano-Nelson (HN)~\cite{NHatano1996} and Aubry-Andr\'e (AA) models~\cite{SAubry1980} to illustrate our ideas. The corresponding Hamiltonian reads
\begin{equation}\label{1DHNAA}
    \begin{aligned}
H=&\sum_{n=1}^{N-1}(J_{L}a_{n}^{\dagger}a_{n+1}+J_{R}a_{n+1}^{\dagger}a_{n})+\sum_{n=1}^{N-1}J_{\mathrm{AA}}(b_{n}^{\dagger}b_{n+1}\\
&+b_{n}b_{n+1}^{\dagger})+\sum_{n=1}^{N}V_{n}b_{n}^{\dagger}b_{n}+t\sum_{n=1}^{N}(a_{n}^{\dagger}b_{n}+b_{n}^{\dagger}a_{n}),
\end{aligned}
\end{equation}
where $a_{n}$, $b_{n}$ ($a_{n}^{\dagger}$, $b_{n}^{\dagger}$) are the annihilation (creation) operators of the $n$-th unit cell in HN and AA chains, respectively. $J_{L/R}=Je^{\pm g}$ and $J_{\mathrm{AA}}$ represent the hopping strength within the HN chain and AA chain, respectively, while $t$ is the coupling strength between the two. $V_{n}=2\lambda\cos(2\pi\alpha n+\theta)$ denotes the quasiperiodic potential, where $\lambda$, $\alpha$, and $\theta$ represent the strength of the potential, an irrational number, and a phase offset, respectively. Since the choice of left and right eigenstates does not qualitatively alter the AS dualism, we select the right eigenstate for all subsequent calculations. Without loss of generality, we set $\theta=0$, $\alpha=(\sqrt{5}-1)/2$ and $J=1$ as a unit energy in numerical calculation. We choose $N=144$ to make $N\alpha$ close to an integer, which is more reasonable under PBCs.

Since the main concern of this Letter is AS dualism, we directly give that the AA chain is in Anderson localization, i.e., the case of $J_{\mathrm{AA}}=0$ (see supplementary materials~\cite{Supplement} for the $J_{\mathrm{AA}}\neq 0$ case). Furthermore, we consider a disordered potential in supplementary materials~\cite{Supplement}, which is qualitatively similar to the quasiperiodic case. We focus on the quasiperiodic potential in the main text since for $J_{\mathrm{AA}}=0$ the Lyapunov exponents (LEs) can be obtained analytically, thus revealing AS dualism.


To demonstrate the characteristics of localization and non-Hermitian topology, we calculate the inverse participation ratio (IPR) and winding number~\cite{ZGong2018} of the HN-AA coupling model~\eqref{1DHNAA}, i.e.,
\begin{equation}\label{1DIPR}
    \begin{aligned}
\xi=\frac{\sum_{n=1}^{N}(|\psi_{a,n}|^4+|\psi_{b,n}|^4)}{\left[\sum_{n=1}^{N}(|\psi_{a,n}|^2+|\psi_{b,n}|^2)\right]^2}
\end{aligned},
\end{equation}
where $\psi_{a/b,~n}$ denote the probability amplitudes, and
\begin{equation}\label{omega}
\omega(E_{b})=\lim_{N\rightarrow\infty}\frac{1}{2\pi i}\int_{0}^{2\pi} d\phi\frac{\partial}{\partial \phi} \ln\left[\mathrm{ det}(H-E_{b})\right],
\end{equation}
where $E_{b}$ is the base point and $\phi$ is the winding flux~\cite{ZGong2018}. Then we have $J_{L}\rightarrow J_{L}e^{i\phi/N}$ and $J_{R}\rightarrow J_{R}e^{-i\phi/N}$.

\begin{figure}[htbp]
	\centering	\includegraphics[width=8.5cm]{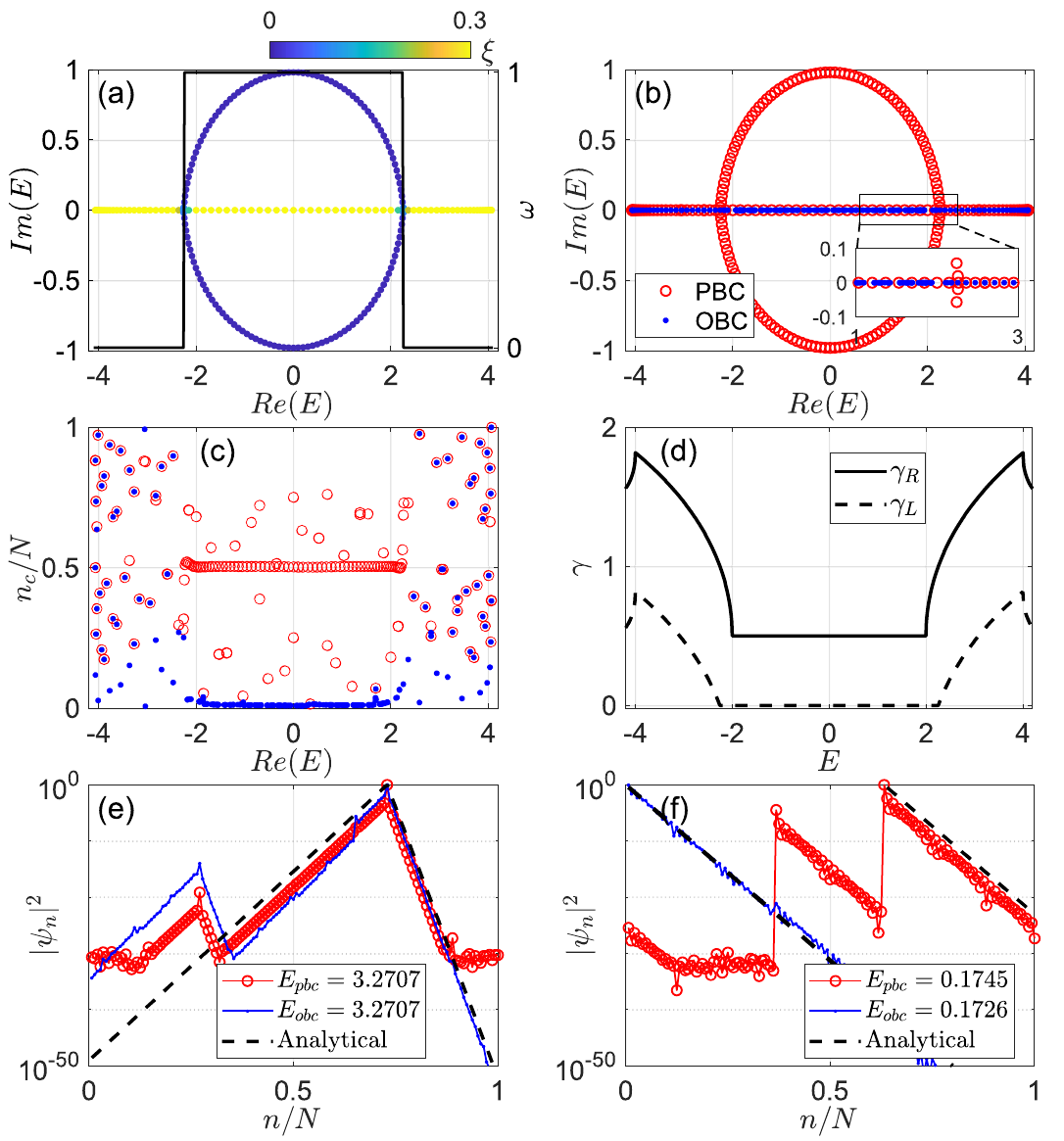}
	\caption{(a) The IPR $\xi$ of all eigenstates in the complex energy plane. Right axis: The corresponding winding number $\omega$ versus $E=Re(E)$. (b) Eigenvalues under different boundary conditions. (c) The positions of center-of-mass for all eigenstates under different boundary conditions. (d) The left and right LEs versus $E$. Probability distributions of eigenstates inside (e) and outside (f) the point gap. Throughout, parameters $t=0.5$, $g=0.5$, $\lambda=2$, and $N=144$. }\label{F2ASD}
\end{figure}

The results show that under PBC, the IPRs corresponding to the eigenstates on the AA chain all exhibit localized state features, while the IPRs corresponding to the eigenstates on the HN chain are manifested as the extended state~(see Fig.~\ref{F2ASD}(a)). In other words, the inter-chain coupling $t$ does not change the Anderson modes of the AA chain under PBC. The corresponding winding number $\omega=1$ ($\omega=0$) in (out of) the point gap, indicating that the corresponding eigenstates will (will not) collapse into skin states under OBC. In concrete terms, under OBC, the corresponding eigenvalues of AA chain within the point-gap-loop region exhibits non-Hermitian topological properties at this time, rather than Anderson localization~(see Fig.~\ref{F2ASD}(b)). This indicates that the eigenstates within the point-gap-loop region have duality dependent on boundary conditions. Since the corresponding wave functions show Anderson modes under PBC while skin modes under OBC, we refer to this phenomenon as Anderson-Skin dualism. 

Note that, from the inset of Fig.~\ref{F2ASD}(b), one can find that the eigenvalues outside the point-gap-loop are robust against the boundary conditions, while the eigenvalues inside are highly sensitive to different boundaries.

Furthermore, in Fig.~\ref{F2ASD}(c), we exhibit the corresponding center of mass’ positions $n_c=\sum_{n=1}^{N}n|\psi_{n}|^{2}$, where $|\psi_{n}|^2$ is defined as $|\psi_{n}|^2=|\psi_{a,n}|^2+|\psi_{b,n}|^2$ of all eigenstates. Specifically, under PBC, the eigenstates’ centers of mass corresponding to AA chain are randomly distributed, which is the evidence of Anderson localization; whereas the eigenstates’ centers of mass corresponding to NH chain are distributed at $n_c=0.5$, indicating that these wave functions are evenly distributed over all sites, i.e., the extended states. On the other hand, under OBC, the eigenstate corresponding to the HN chain will change from the extended state to the skin state, which is consistent with previous studies~\cite{SZLi2024,YLiu2021}. However, counterintuitively, the Anderson modes of the AA chain originally inside the point-gap-loop also become skin states under OBC, while the states outside the loop remain unchanged.


Another very effective indicator for distinguishing skin modes from Anderson modes is the Lyapunov exponent. HN-AA coupled ladder~\eqref{1DHNAA}, as an non-reciprocal system, can be analytically studied by the scheme of the asymmetric transfer matrix analysis~\cite{SZLi2024}. One can get two distinct LEs for a common non-reciprocal Anderson mode, whereas only one-side LE for a skin mode. The eigenequation of Hamiltonian~\eqref{1DHNAA} reads
\begin{equation}
\begin{aligned}
&J_{R}\psi_{a,n-1}+t\psi_{b,n}+J_{L}\psi_{a,n+1}=E\psi_{a,n},\\
&t\psi_{a,n}+V_{n}\psi_{b,n}=E\psi_{b,n}.
\end{aligned}
\end{equation}
By eliminating $\psi_{b,n}$, we obtain
\begin{equation}\label{SimpEq}
\begin{aligned}
&J_{R}\psi_{a,n-1}+\frac{t^2}{E-V_{n}}\psi_{a,n}+J_{L}\psi_{a,n+1}=E\psi_{a,n}.
\end{aligned}
\end{equation}
Then one can get the corresponding LEs, i.e.,
\begin{equation}\label{gamma_main}
\gamma_{L/R}=\max\left\{\gamma\pm g,0\right\},
\end{equation}
where
\begin{equation}
\gamma=\left\{\begin{matrix}
\ln\left|\dfrac{\lambda(E+\sqrt{E^2-4})}{E+\sqrt{E^2-4\lambda^2}}\right|,  & |E|\ge 2\lambda~\&~|E^2|\ge 4, \\
\ln\left|\dfrac{E+\sqrt{E^2-4}}{2}\right|,  & |E|< 2\lambda~\&~|E^2|\ge 4,\\
\ln\left|\dfrac{2\lambda}{(E+\sqrt{E^2-4\lambda^2})}\right|,  & |E|\ge 2\lambda~\&~|E^2|< 4,\\
0,  & |E|< 2\lambda~\&~|E^2|< 4.\\
\end{matrix}\right.
\end{equation}

In Fig.~\ref{F2ASD}(d), we exhibit LEs versus $E$. For the states outside the point gap ($|E|>J_{R}+J_{L}$), there are two LEs on both sides of the localized center, which can be fitted by~\cite{HJiang2019}
\begin{equation}\label{fit}    
|\psi_{n}|\propto \left\{\begin{matrix}e^{-\gamma_{R}(n-n_{0})},& n>n_{0}, \\e^{-\gamma_{L}(n_{0}-n)},& n<n_{0},\end{matrix}\right.
\end{equation}
where $n_0$ is the center of localization. This implies that the eigenstates within this region are all common Anderson modes. 

On the other hand, for the eigenstates inside the point-gap-loop ($|E|<J_{R}+J_{L}$), only one side LE left, i.e., $\gamma_{R}\neq 0$ and $\gamma_{L}= 0$, which is the evidence of skin modes (see Fig.~\ref{F2ASD}). The corresponding LE can be fitted by $|\psi_{n}|\propto e^{-\gamma_R(n-n_0)}$. 

To visually show AS dualism, we plot the density distributions of two typical states under PBC and OBC, so as to compare them with the analytical solutions of LEs. It turns out that the analytical solutions of LE are in complete agreement with the previous analysis of IPR, winding number, center-of-mass position, etc. Note that, unlike the conventional common Anderson modes that all have two-side LEs~\cite{HJiang2019,SZLi2024,YLiu2021}, AS dualism state has Anderson modes with only one-side LE localized in the bulk under PBC. Moreover, the LE of these unconventional Anderson modes is the same as that of the skin modes under OBC. In other words, although the AS dualism mode exhibits different characteristics under different boundary conditions, the wave function structures of the two are the same. Similarly, in both 2D and 3D cases, we can also observe such interesting AS dualism modes (see SM~\cite{Supplement} for details).

\textcolor{blue}{\emph{Occupation-denpendent AS dualism of interacting systems.}}---Now, we turn to discuss AS dualism in many-body systems. We focus on a simple case involving two hard-core bosons loaded in a 1D HN-AA ladder lattice. By introducing inter-chain interaction $H_{int} =U\sum_{n=1}^{N}a_{n}^{\dagger}a_{n}b_{n}^{\dagger}b_{n}$ into Eq.~\eqref{1DHNAA}, one can obtain the total Hamiltonian $\tilde{H}=H+H_{int}$~\cite{YQin2024}. 

\begin{figure}[htbp]
	\centering	\includegraphics[width=8.5cm]{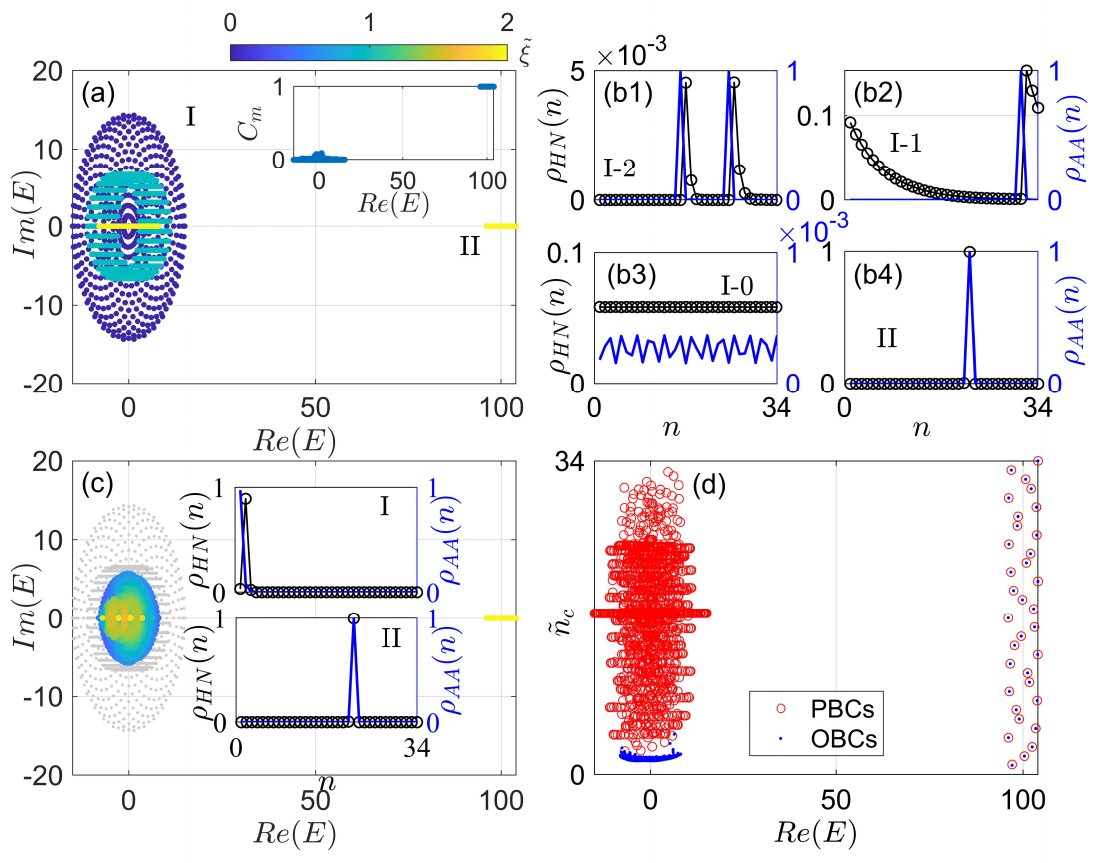}
	\caption{Many-body IPRs of all eigenstates with PBCs (a) and OBCs (c), where the gray dots represent the eigenvalues under PBCs. The inset illustrates the sublattice correlation $C_{m}$ versus $Re(E)$. Particle distributions for $E=-0.7522$ (b1), $-2.9020+4.5682i$ (b2), $E=-2.8849+6.5858i$(b3), and $E=96.7366$ (b4). I-$\mu$ means that $\mu$ particles are occupied on the AA chain. The inset of (c) shows the OBC particle distribution for the eigenstates in cluster I and II, respectively. (d) The mean positions $\tilde{n}_c$ versus Re($E$). Thourghout, $g=2$, $\lambda=2$, $t=0.5$, $U=100$. $N=34$.}\label{F3}
\end{figure}

To describe the localization properties of the many-body eigenstates, we define many-body IPR as 
\begin{equation}
\tilde{\xi}=\sum_{n}\sum_{s=\mathrm{HN,AA}}|\rho_{s}(n)|^2,   
\end{equation}
where $\rho_{\mathrm{HN}}(n)=\left<\psi\right|a_{n}^{\dagger}a_{n}\left|\psi\right>$ and $\rho_{\mathrm{AA}}(n) = \left<\psi\right|b_{n}^{\dagger}b_{n}\left|\psi\right>$ represent the particles' occupancy probabilities on the HN chain and the AA chain, respectively, and  $\left|\psi\right>$ denotes the normalized eigenstate of many-body Hamiltonian $\tilde{H}$. The corresponding many-body IPRs are shown in Fig.~\ref{F3}(a), which indicates that the eigenspectrum is divided into two clusters in the complex plane. 

Cluster I has a complex spectral structure (multi-looped point gaps), which contains the eigenstates of $\tilde{\xi}=0,~1,~2$, corresponding to $0,~1$ and $2$ particles’ localization on the AA chain, respectively. There will be the phenomenon of AS dualism in this case. However, Cluster II only has pure real eigenvalues, which fall within the range of $U-2\lambda<E<U+2\lambda$, and $\tilde{\xi}=2$ always. In this case no AS dualism will emerge. This is because different hard-core bosons cannot occupy the same site. Therefore, the inter-chain coupling is suppressed in the double-occupation configuration, which eventually leads to the highly localized feature of the wave function. Different from the AS dualism of the single-particle, the many-body version of the AS dualism phenomenon has an occupation-dependent characteristic.

The inset of Fig.~\ref{F3}(a) exhibits the corresponding sublattice correlation function
\begin{equation}
C_{m}=\sum_{n=1}^{N}\left<\psi_{m}\right|a_{n}^{\dagger}a_{n}b_{n}^{\dagger}b_{n}\left|\psi_{m}\right>, 
\end{equation}
where $m$ is the level index. The results reveal that the correlation corresponding to cluster I tends to zero, i.e., $C_{m} \sim 0$, which indicates that cluster I is composed of single-occupation states. On the other hand, cluster II corresponds to the case of $C_{m}\sim1$, which is the evidence of a double-occupation state. The sublattice correlation function once again demonstrates that AS dualism has an occupation-dependent characteristic.

In addition, under PBC, we show the particle distribution corresponding to the eigenstates in regions I and II (see Fig.~\ref{F3}(b1)-(b4)). Depending on the number of particles $\mu$ occupying the AA chain, one can further divide the states of cluster I into three different cases. Specifically, for the $\mu=2$ ($\tilde{\xi}\sim 2$) case, although there is a small probability of penetration into the HN chain, the two particles are mainly localized on the AA chain, and the system mainly exhibits the characteristics of Anderson mode (see Fig.~\ref{F3}(b1)). For the $\mu=1$ ($\tilde{\xi}\sim 1$) case, there is one particle in each of the two chains. The particle in the AA chain is localized, while the particle in the HN chain is bound to the sites near the particle in the AA chain in an asymmetric localized structure due to the influence of AA chain. For $\mu=0$ ($\tilde{\xi}\sim0$), i.e., both particles are placed on the HN chain, the system shows the characteristics of the HN chain, namely the extended state (see Fig.~\ref{F3}(b3)). At last, as analyzed earlier, the eigenstates in cluster II have both particles completely localized on the same cell, forming a stable double-occupation localized state. This also comfirms that for the many-body system, only the non-zero single-occupation situation on the AA chain will give rise to AS dualism (see Fig.~\ref{F3}(b1)(b2)). This once again demonstrates that AS dualism in many-body systems is occupation-dependent.

We further discuss the properties of the eigenstates corresponding to Clusters I and II under OBC. Fig.~\ref{F3}(c) shows the spectral structure in the complex plane and many-body IPR corresponding to all eigenstates under OBC. The results show that the spectrum of Cluster I will collapse, indicating the existence of the skin effect (see insets of (c)), while Cluster II is independent on the boundary conditions, which means that these states are stable Anderson modes. Finally, we exhibit the corresponding mean position of each eigenstate, i.e.,
\begin{equation}
\tilde{n}_{c}=\frac{1}{2}\sum_{n}n[\rho_{\mathrm{HN}}(n)+\rho_{\mathrm{AA}}(n)].    
\end{equation}
By comparing what happens under PBC and OBC, one can find that the distribution of particles in Clusters I is located at the bulk (boundary) under PBC (OBC), while the center-of-mass position in Clusters II does not change with the boundary conditions (see Fig.~\ref{F3}(d)). This once again confirms our conclusion: The localized states within the point gaps (the region of Cluster I) have AS dualism.

\textcolor{blue}{\emph{Conclusion.}}---We uncover a new localization phenomenon caused by the interplay of point gap topology and disorder, which we dub ``AS dualism". The AS dualism state is manifested as bulk localization (Anderson modes) under PBC, whereas boundary localization (Skin modes) under OBC. Firstly, we study the 1D case through exact solutions and numerical simulations, confirming that AS dualism will emerge in the HN and AA coupling chains. The formalism of AS dualism relies only on the interference between point-gap topology and Anderson localization on different subsystems, which can both be robust in certain higher-dimensional models~\cite{KZhang2022}. In our supplementary 
materials, we provide explicit examples in reciprocal, 2D, 3D~\cite{Supplement} cases, demonstrating the universality of the AS dualism. At last, we show that the AS dualism of many-body systems will be affected by the particle occupancy, i.e., it is occupation-dependent. Since numerous experimental platforms have been developed to realize topological point gaps and skin effects, such as ultracold atoms~\cite{WGuo2020,QLiang2022}, photonic systems~\cite{ZFang2022,WWang2023}, and topological circuits~\cite{LSu2023,THelbing2020,XZhang2021,CXGuo2024,LLi2025}, we have confidence to expect that the AS dualism will soon be observed in tabletop experiments.

\textcolor{blue}{\emph{Note added:}} A recent paper on non-Hermitian quasicrystal systems Arxiv:2412.04344~\cite{SPadhi2024} shows an anomalous spectral structure similar to that of Fig.~\ref{F2ASD}(a). Our work demonstrates that this spectral behavior originates from AS dualism in the dual space.

\textcolor{blue}{\emph{Acknowledgements.}}---This work was supported by the National Key Research and Development Program of China (Grant No.2022YFA1405300),  Innovation Program for Quantum Science and Technology (Grant No. 2021ZD0301700),  the Guangdong Basic and Applied Basic Research Foundation (Grant No.2021A1515012350), Guangdong Provincial Quantum Science Strategic Initiative(Grants No. GDZX2304002 and GDZX2404001), and the Open Fund of Key Laboratory of Atomic and Subatomic Structure and Quantum Control (Ministry of Education).

\emph{Data availability}.---The data that support the findings of this article are not publicly available. The data are available from the authors upon reasonable request.

\global\long\def\id{\mathbbm{1}}
\global\long\def\ui{\mathbbm{i}}
\global\long\def\ud{\mathrm{d}}

\setcounter{equation}{0} \setcounter{figure}{0}
\setcounter{table}{0} 
\renewcommand{\theparagraph}{\bf}
\renewcommand{\thefigure}{S\arabic{figure}}
\renewcommand{\theequation}{S\arabic{equation}}

\onecolumngrid
\flushbottom

\newpage
\section*{Supplementary materials for: ``Anderson-skin dualism: A boundary-dependent effect in non-Hermitian disordered coupled systems''}

\section{I. The case of random disorder}
Here, we consider a Hatano-Nelson chain coupled to a one-dimensional Anderson-localized chain, with the Hamiltonian written as
\begin{equation}\label{1DHNrandom}
H=\sum_{n=1}^{N-1}(J_{L}a_{n}^{\dagger}a_{n+1}+J_{R}a_{n+1}^{\dagger}a_{n})+\sum_{n=1}^{N-1}J_{\mathrm{rand}}(b_{n}^{\dagger}b_{n+1}+b_{n}b_{n+1}^{\dagger})+\sum_{n=1}^{N}W_{n}b_{n}^{\dagger}b_{n}+t\sum_{n=1}^{N}(a_{n}^{\dagger}b_{n}+b_{n}^{\dagger}a_{n}),
\end{equation}
where \(W_{n}\) is drawn uniformly from \([-W, W]\). In the reciprocal case, unlike under quasiperiodic potentials, arbitrarily small uncorrelated random disorder can localize a 1D system. In the non-reciprocal case, however, the system exhibits two Lyapunov exponents (LEs), with one typically increasing while the other decreasing. This implies that a localization transition occurs only when both LEs are positive. Consequently, even in the presence of random disorder, non-Hermitian 1D systems can support extended states~\cite{ZGong2018}.

 \begin{figure}[htbp]
	\centering	\includegraphics[width=13cm]{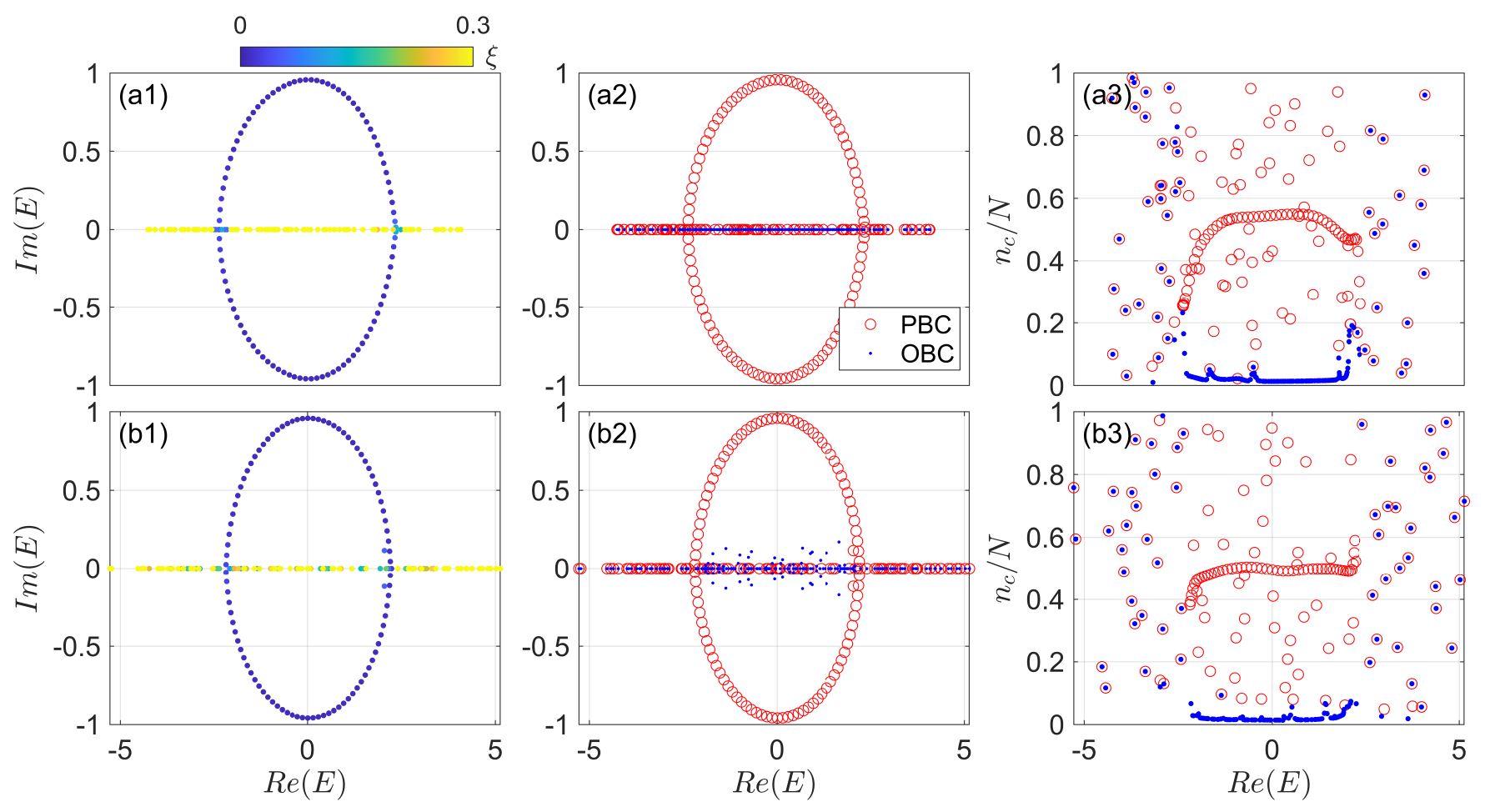}
	\caption{Under the condition of $J_{\mathrm{rand}}=0$ [$J_{\mathrm{rand}}=1$], the IPR (a1) [(b1)] corresponding to the eigenvalues under PBC, as well as the comparison of eigenstates (a2) [(b2)] and the positions of the center-of-mass (a3) [(b3)] with different boundary conditions. Throughout, $W=4$, $t=0.5$, $J=1$, $g=0.5$, and $N=100$. } \label{disorder}
\end{figure}
Fig.~\ref{1DHNrandom} illustrates the IPR in the complex energy plane, the eigenvalues under different boundary conditions, and the center-of-mass positions of all eigenstates. We see that for both $J_{\mathrm{rand}}=0$ [(a1)-(a3)] and $J_{\mathrm{rand}}=1$ [(b1)-(b3)], the system exhibits AS dualism, where Anderson-localized states within the point gap under PBCs manifest as skin states under OBCs. However, under PBCs, some extended states deviate from the relation $n_{c}/N \sim 0.5$, which we attribute to disorder-induced inhomogeneity-a phenomenon also observed in non-reciprocal Anderson models.

Thus, AS dualism depends not on the detailed form of the disordered potential but only on the existence of localized states inside the point gap.

\section{II. The derivation details of the Lyapunov exponent}
The eigenequation for $H$ in main text is 
\begin{equation}
\label{S1}
\begin{aligned}
&J_{R}\psi_{a,n-1}+t\psi_{b,n}+J_{L}\psi_{a,n+1}=E\psi_{a,n},\\
&t\psi_{a,n}+V_{n}\psi_{b,n}=E\psi_{b,n}.
\end{aligned}
\end{equation}
One can find that there is a relationship between $\psi_{a,n}$ and $\psi_{b,n}$, namely $\psi_{a,n}=(V_{n}-E)\psi_{b,n}/t$. Substituting this into Eq.~\eqref{S1}, one can obtain a new equation for the component $\psi_{a}$, i.e.,
\begin{equation}\label{EqH}
\begin{aligned}
&J_{R}\psi_{a,n-1}+\frac{t^2}{E-V_{n}}\psi_{a,n}+J_{L}\psi_{a,n+1}=E\psi_{a,n}.
\end{aligned}
\end{equation}
This eigenvalue equation is analogous to that of the nonreciprocal AA model. Therefore, one can employ the asymmetric transfer matrix method to determine the Lyapunov exponents of the eigenstates in both leftward and rightward directions~\cite{SZLi2024}. In concrete terms, one can obtain the corresponding transfer matrix from left $(\psi_{a,n}, \psi_{a,n-1})^{T}$ to right $(\psi_{a,n+1}, \psi_{a,n})^{T}$ as
\begin{equation}\label{TLAAH}
\begin{pmatrix}
\psi_{a,n+1} 
\psi_{a,n}
\end{pmatrix}=T_{L,n}
\begin{pmatrix}
\psi_{a,n} \\
\psi_{a,n-1}
\end{pmatrix},\\ ~\text{where}~T_{L,n}=\begin{pmatrix}
\frac{E(E-V_{n})-t^2}{J_{L}(E-V_n)}  & -\frac{J_{R}}{J_{L}}\\
 1 &0
\end{pmatrix},
\end{equation}
as well as the transfer matrix from right $(\psi_{a,n+1}, \psi_{a,n})^{T}$ to left $(\psi_{a,n-1}, \psi_{a,n})^{T}$ as
\begin{equation}\label{TR}
\begin{pmatrix}
\psi_{a,n-1} \\
\psi_{a,n}
\end{pmatrix}=T_{R,n}
\begin{pmatrix}
\psi_{a,n} \\
\psi_{a,n+1}
\end{pmatrix},~\text{where}~T_{R,n}=
\begin{pmatrix}
\frac{E(E-V_{n})-t^2}{J_{R}(E-V_n)}  & -\frac{J_{L}}{J_{R}}\\
 1 &0
\end{pmatrix}.
\end{equation}
The corresponding LEs read
\begin{equation}
\begin{aligned}
&\gamma_{L}(E)=\lim_{L\rightarrow\infty}\frac{1}{N}\ln\left \| \prod_{n=1}^{N} T_{L,n} \right \|,\\
&\gamma_{R}(E)=\lim_{L\rightarrow\infty}\frac{1}{N}\ln\left \| \prod_{n=1}^{N} T_{R,n} \right \|,
\end{aligned}
\end{equation}
where $\prod_{n=1}^{N} T_{L,n}=T_{L,N}\dots T_{L,2}T_{L,1}$, and $\prod_{n=N}^{1} T_{R,n}=T_{R,1}\dots T_{R,N-1}T_{R,N}$. The symbol $\left \| \cdot  \right \| $ denotes the norm of $\prod_{n=1}^{N} T_{L,n}$ (or $\prod_{n=N}^{1} T_{R,n}$), which in fact takes the absolute value of the largest eigenvalue in the matrix. Then, one can get Eq.~\eqref{TLAAH} and \eqref{TR} in the form 
\begin{equation}
T_{L/R,n}=\frac{1}{J_{L/R}(E-V_{n})}\begin{pmatrix}
E^2-EV_{n}-t^2  & -J_{R/L}(E-V_{n}) \\
J_{L/R}(E-V_{n}) &0
\end{pmatrix},
\end{equation}
so that the LE of the system is $\gamma_{L/R}(E)=\gamma_{L/R,A}(E)+\gamma_{L/R,B}(E)$ consisting of two parts, where $\gamma_{L/R,A}(E)$ is given by~\cite{SLonghi2019}
\begin{equation}\label{gammaA}
\gamma_{L/R,A}(E)=\lim_{N\rightarrow \infty}\frac{1}{N}\ln\left \|\prod_{j=1}^{L}\frac{1}{J_{L/R}(E-V_{n})}   \right \|=\left\{
\begin{matrix}
\ln\left|\frac{2}{J_{L/R}(E+\sqrt{E^2-4\lambda^2})}\right|,  &|E|\ge 2\lambda, \\
\ln\left|\frac{1}{J_{L/R}\lambda}\right|, & |E|< 2\lambda.
\end{matrix}\right.
\end{equation}
For the $\gamma_{L/R,B}(E)$, we employ the Avila's global theory~\cite{Avila2015}, which results in
\begin{equation}
\gamma_{L/R,B}(E)=\left\{\begin{matrix}
\ln\left|\frac{\lambda E+\lambda\sqrt{E^2-4}}{2}\right|,  & |E|\ge 2, \\
\ln|\lambda|,  & |E|<2.
\end{matrix}\right.
\end{equation}
Therefore, the total LE is
\begin{equation}
\gamma_{L/R}(E)=\left\{\begin{matrix}
\ln\left|\dfrac{\lambda(E+\sqrt{E^2-4})}{J_{L/R}(E+\sqrt{E^2-4\lambda^2})}\right|,  & |E|\ge 2\lambda~\&~|E|\ge 2, \\
\ln\left|\dfrac{E+\sqrt{E^2-4}}{2J_{L/R}}\right|,  & |E|< 2\lambda~\&~|E|\ge 2,\\
\ln\left|\dfrac{2\lambda}{J_{L/R}(E+\sqrt{E^2-4\lambda^2})}\right|,  & |E|\ge 2\lambda~\&~|E|< 2,\\
\ln\left|\dfrac{1}{J_{L/R}}\right|,  & |E|< 2\lambda~\&~|E|< 2,\\
\end{matrix}\right.
\end{equation}
Note that, since Eq.~\eqref{gammaA} is solved only for real $E$, our results describe solely eigenstates with real eigenvalues. In the main text, we find that the eigenvalues of the localized and skin states are real, which is sufficient for the discussion of AS dualism.

\section{III. Anderson-Skin dualism for case \texorpdfstring{$J_{\mathrm{AA}}\neq 0$}{J_AA not equal to 0}}

\begin{figure}[htbp]
	\centering	\includegraphics[width=13cm]{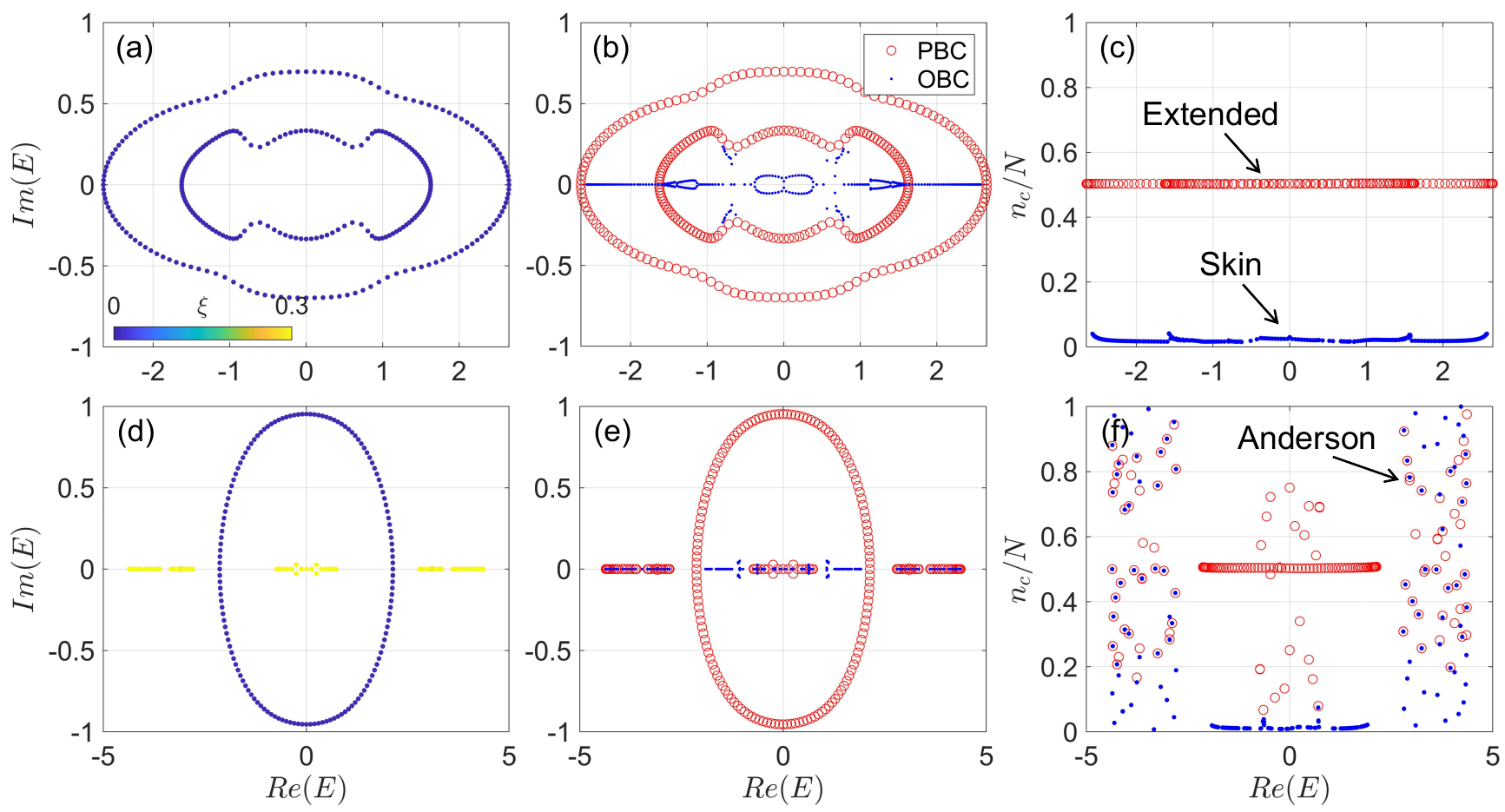}
	\caption{Under the condition of $\lambda=0.2$ [$\lambda=2$], the IPR (a) [(d)] corresponding to the eigenvalues under PBC, as well as the comparison of eigenstates (b) [(e)] and the positions of the center-of-mass (c) [(f)] with different boundary conditions. Throughout, $t=0.5$, $J=1$, $g=0.5$, and $N=144$. } \label{J1}
\end{figure}

Here, we take the $J_{\mathrm{AA}}=1$ case as an example to discuss the more general case of $J_{\mathrm{AA}}\neq 0$. From the Hamiltonian $H$ in main text, one can find that the critical point of Anderson transition of the AA chain is $\lambda=J_{\mathrm{AA}}$~\cite{SAubry1980}. In Fig.~\ref{J1}(a)-(c) [(d)-(f)], we exhibit the results for the AA chain with $\lambda=0.2$ [$\lambda=2$], i.e., the extended [localized] phase.

On the one hand, for the extended phase ($\lambda=0.2$), the system exhibits separated spectra in the complex plane, forming two closed point gaps. Under OBCs, the spectrum collapses, indicating the presence of skin effect (see Fig.~\ref{J1}(b)). Fig.~\ref{J1}(c) shows positions of the center-of-mass for the eigenstates under different boundary conditions. For OBCs, all eigenstates are localized at the boundary, whereas for PBCs, the eigenstates are uniformly distributed, resulting in a center located at the middle of the chain. This means that there is no AS dualism when the AA chain is extended.

On the other hand, for the localized phase ($\lambda=2$), the results are consistent with the discussion in the main text: The eigenstates in the point gaps have the characteristics of AS dualism [see Fig.~\ref{J1}(d)-(f)]. In contrast, localized states outside the point gaps are always localized in the bulk.

\section{IV. Anderson-Skin dualism in reciprocal systems}
 \begin{figure}[htbp]
	\centering	\includegraphics[width=7cm]{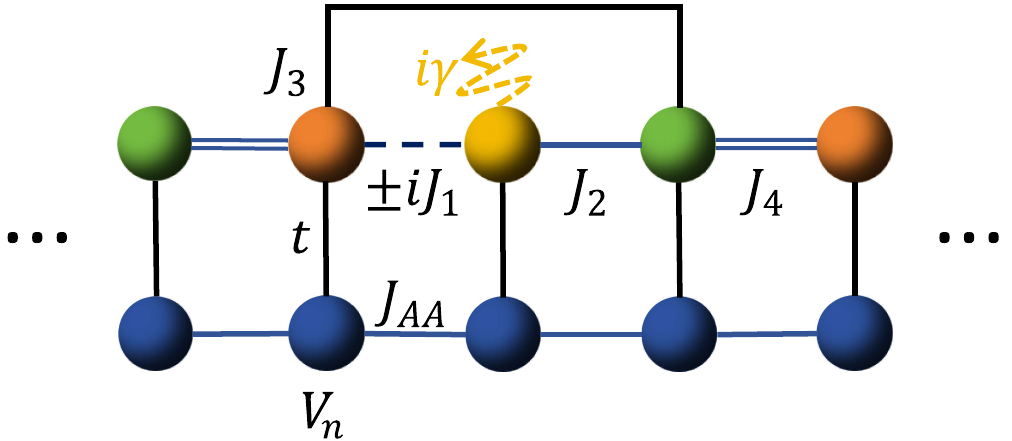}
	\caption{This schematic diagram of the coupled AA-AB model illustrates reciprocal AS dualism.} \label{Fs5}
\end{figure}

The skin effect can arise not only from non-reciprocal hopping but also from reciprocal dissipative systems. This implies that AS dualism can occur within reciprocal systems. Here we replace the chain that provides the topological point gap by a dissipative Aharonov-Bohm (AB) chain~\cite{QLiang2022}, while retaining the AA-model chain for the localized states, as shown in Fig.~\ref{Fs5}. Thus, the total Hamiltonian is written as $H=H_{\mathrm{AB}}+H_{\mathrm{AA}}+H_{\mathrm{c}}$, where
\begin{equation}\label{1Dreciprocal}
\begin{aligned}
&H_{\mathrm{AB} }  = \sum_{n  = 1}(iJ_{1}a_{n}^{\dagger}b_{n}+J_{2}b_{n}^{\dagger}c_{n}+J_{3}a_{n}^{\dagger}c_{n}+J_{4}a_{n+1}^{\dagger}c_{n}+\mathrm{H.c.} )+\sum_{n=1}(-i\gamma)b_{n}^{\dagger}b_{n},\\
&H_{\mathrm{AA} }  = \sum_{n  = 1}J_{\mathrm{AA}}(d_{n}^{\dagger}d_{n+1}+d_{n+1}^{\dagger}d_{n})+\sum_{n  = 1}V_{n}d_{n}^{\dagger}d_{n},\\
&H_{c}=\sum_{n=1}t(d_{3n-2}^{\dagger}a_{n}+d_{3n-1}^{\dagger}b_{n}+d_{3n}^{\dagger}c_{n})+\mathrm{H.c.} .
\end{aligned}
\end{equation}
The Hamiltonian $H_{AB}$ consists of $N/3$ primitive cells, each containing three sublattice sites $a$, $b$, and $c$. In $H_{AB}$, the operators $a_{n}^{\dagger}$ ($a_{n}$), $b_{n}^{\dagger}$ ($b_{n}$), and $c_{n}^{\dagger}$ ($c_{n}$) are the creation (annihilation) operators for the $a$, $b$, and $c$ sites of the $n$-th cell, respectively. The intracell hopping is $\pm iJ_{1}$ (between sites $a$ and $b$), $J_{2}$ (between $a$ and $c$), and $J_3$ (between $b$ and $c$). The intercell hopping strength from $a$ to $c$ is $J_{4}$. $H_{AA}$ represents the AA chain, where we take $V_{n}=2\lambda\cos(2\pi\alpha n)$. The coupling strength between the AA and AB chains is $t$. Note that, because the AB chain has three sublattices per cell, the $a$, $b$, and $c$ sites of cell $n$ coupled to sites $3n-2$, $3n-1$, and $3n$ of the AA chain, respectively. 
 \begin{figure}[htbp]
	\centering	\includegraphics[width=15cm]{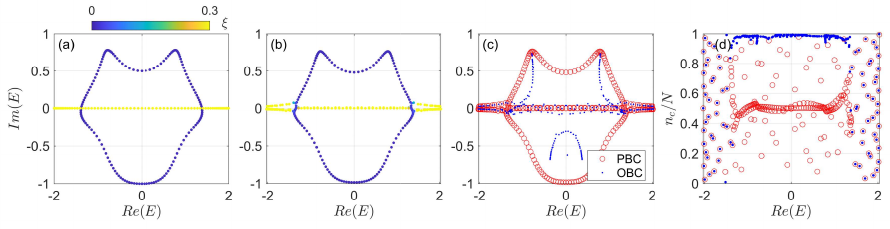}
	\caption{ Under the condition of $t=0$ [$t=0.2$], the IPR (a) [(b)] corresponding to the eigenvalues under PBC. (c) eigenvalues under different boundary conditions. (d) The positions of the center-of-mass with different boundary conditions. Throughout, $\lambda=1$, $J_{1}=1$, $J_{2}=1.5$, $J_{3}=0.5$, $J_{4}=0.5$, $\gamma=3$ and $N=144$. } \label{Fs6}
\end{figure}

When the interchain coupling vanishes ($t=0$), the AB-chain spectrum ($\xi \sim 0$) exhibits a topological point gap in the complex energy plane [Fig.~\ref{Fs6}(a)]. When the interchain coupling is finite ($t=0.2$), the spectrum shifts slightly [Fig.~\ref{Fs6}(b)], while the localized states remain inside the point gap. Figs.~\ref{Fs6}(c,d) compare the eigenvalues and the center-of-mass positions of all eigenstates under different boundary conditions. Similar to the non-reciprocal case, localized states lying inside the point gap exhibit AS dualism, whereas those outside the point gap are unaffected by the boundary conditions.

\section{V. Anderson-Skin dualism in high dimensional cases}
\subsection{A. case of 2D }
In this subsection, we prove that the AS dualism can emerge in 2D NH-AA coupled systems. The corresponding Hamiltonian reads
\begin{equation}
    H = H_{\mathrm{HN}}^{2D} + H_{\mathrm{AA}}^{2D} + H_C
\end{equation}
with
\begin{equation}
\begin{aligned}
&H_{\mathrm{HN}}^{2D} = \sum_{x,y}^{N_x,N_y} \left( J_{R,x} a_{x+1,y}^{\dagger} + J_{L,x} a_{x-1,y}^{\dagger} \right) a_{x,y} + \sum_{x,y=1}^{N_x,N_y} \left( J_{R,y} a_{x,y+1}^{\dagger} + J_{L,y} a_{x,y-1}^{\dagger} \right) a_{x,y}, \\
&H_{\mathrm{AA}}^{2D}= \sum_{x,y}^{N_x,N_y} 2\lambda \left[ \cos(2\pi \alpha_x x + \theta_x) + \cos(2\pi \alpha_y y + \theta_y) \right] b_{x,y}^{\dagger} b_{x,y},~~H_C= \sum_{x,y}^{N_x,N_y} t a_{x,y}^{\dagger} b_{x,y} + \text{H.c.},
\end{aligned}
\end{equation}
where $a_{x,y}$, $b_{x,y}$ ($a_{x,y}^{\dagger}$, $b_{x,y}^{\dagger}$) are the annihilation (creation) operators of the $\{x,y\}$-th unit cell in 2D HN and AA lattices, respectively. $J_{L/R,~x/y}$ represent the hopping strength of the HN chain in the $x/y$ direction to the left/right, while $t$ is the coupling strength between the two. $N_{x/y}$ is the total unit cells in the $x/y$-direction. We set $J_{R,x}=J_{R,y}=Je^{g}$ and $J_{L,x}=J_{L,y}=Je^{-g}$, so that when $g>0$, the skin states are accumulated towards the end of the $x$, $y$-direction.

\begin{figure*}[bh]
	\centering	\includegraphics[width=15cm]{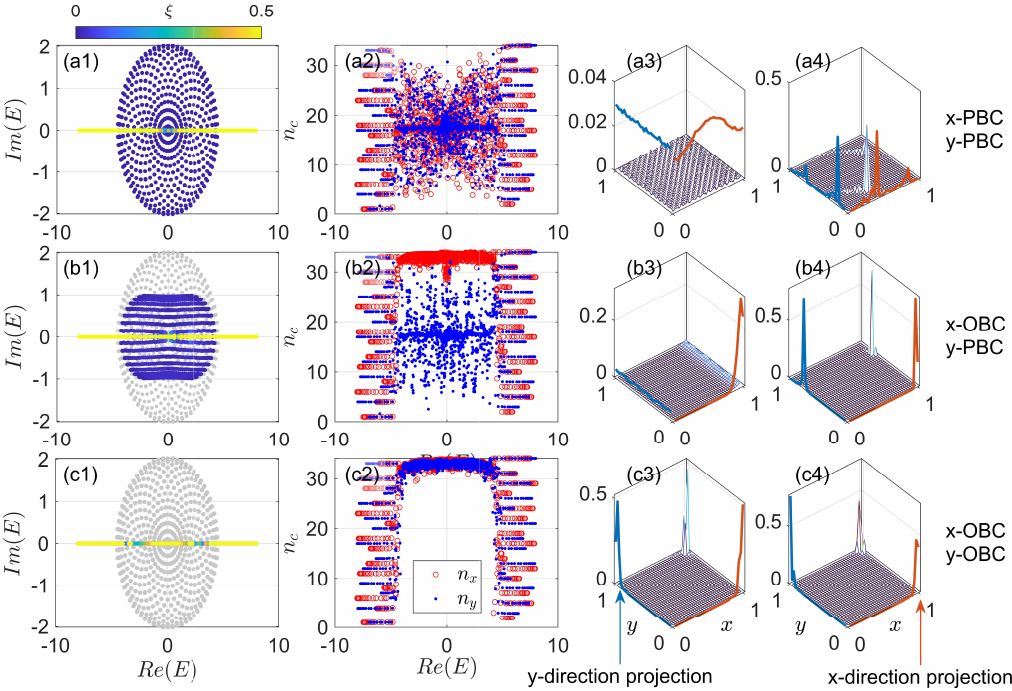}
	\caption{The IPR $\xi$ of all eigenstates in the complex \textcolor{blue}{energy} plane under $xy-PBCs$ (a1), $x-OBC~y-PBC$ (b1), and $xy-OBCs$, respectively. (a2)-(c2) The positions of center-of-mass for all eigenstates as a function of $Re(E)$. (a3)-(c3) and (a4)-(c4) exhibit the probability distributions of two representative eigenstates within the point gap. The eigenvalues for (a3)-(c3) are $E=-0.3038+0.6734$, $E=-0.2689+0.6294$, and $E=-0.307$, while those for (a4)-(c4) are $E=-0.2928$, $E=-0.2746$, and $E=0.1293$, respectively. The other parameters are set as $t=0.5$, $\alpha_{x}=\sqrt{2}$, $\alpha_{y}=(\sqrt{5}-1)/2$, $g=0.5$, $\lambda=2$, and $N_x=N_y=34$.}\label{F2}
\end{figure*}

In Fig.~\ref{F2}, we present the results for the IPR, the positions of eigenstates' center-of-mass, and the distribution of eigenstates under different boundary conditions.

From  Fig.~\ref{F2}(a1)(b1)(c1), under PBCs in both $x$- and $y$-directions, the localized states can exist within the 2D point gap, which is similar to 1D case (see Fig.~\ref{F2}(a1)). However, under OBC in the $x$-direction and PBC in the $y$-direction, the spectrum partially collapses, while still enclosing certain regions, indicating the presence of an $x$-direction skin state (see Fig.~\ref{F2}(b1)). Finally, under OBCs in both $x$- and $y$-directions, all the eigenvalues collapse further onto the real axis (see Fig.~\ref{F2}(c1)). This progressive collapse of the spectrum with changing boundary conditions highlights the existence of a skin state in both directions.

Fig.~\ref{F2}(a2)(b2)(c2) present the eigenstates' positions of center-of-mass in $x$- and $y$-directions under different boundary conditions, which is similar to the results in 1D case. For eigenstates within the point gap, both extended [Fig.~\ref{F2}(a3)] and localized [Fig.~\ref{F2}(a4)] states coexist under PBCs in both directions. Under the case of OBC in $x$-direction and PBC in $y$-direction, the position of center-of-mass of each eigenstate shifts ($n_x \rightarrow N_x$), while $n_y$ remains unchanged. This shift reveals two distinct skin modes. The first, arising from the collapse of states within the point gap, manifests as a first-order skin effect [Fig.~\ref{F2}(b3)]. The second is a hybrid Anderson-skin (AS) mode: it originates from Anderson-localized states in the point gap and behaves as a skin mode in $x$-direction but remains Anderson localized in $y$-direction [Fig.~\ref{F2}(b4)]. This hybrid state indicates that the AS dualism can emerge in only one direction within 2D systems. Furthermore, under OBCs in both the $x$- and $y$-directions, the eigenstates within the point gap consistently transform into corner skin states [Fig.~\ref{F2}(c3)-(c4)]. These results demonstrate richer AS dualism in higher dimensions. In other words, by adjusting the boundary conditions, one can manipulate a localized state to first transform from a localized Anderson mode to a hybrid Anderson-skin mode, and then into a pure skin mode. In contrast, eigenstates outside the point gap characterized by large IPR always correspond to localized Anderson modes, regardless of boundary conditions.

\subsection{B. case of 3D}
Now, let's discuss the 3D case. The corresponding Hamiltonian reads
\begin{equation}
    H = H_{\mathrm{HN}}^{3D} + H_{\mathrm{AA}}^{3D} + H_C
\end{equation}
with
\begin{equation}
\begin{aligned}
H_{\mathrm{HN}}^{3D} = &\sum_{x,y,z=1}^{N_x,N_y,N_{z}} \left( J_{R,x} a_{x+1,y,z}^{\dagger} + J_{L,x} a_{x-1,y,z}^{\dagger} \right) a_{x,y,z} + \sum_{x,y,z=1}^{N_x,N_y,N_{z}} \left( J_{R,y} a_{x,y+1,z}^{\dagger} + J_{L,y} a_{x,y-1,z}^{\dagger} \right) a_{x,y,z}\\&+ \sum_{x,y,z=1}^{N_x,N_y,N_{z}} \left( J_{R,z} a_{x,y,z+1}^{\dagger} + J_{L,z} a_{x,y,z-1}^{\dagger} \right) a_{x,y,z}, \\
H_{\mathrm{AA}}^{2D}=& \sum_{x,y,z=1}^{N_x,N_y,N_{z}} 2\lambda \left[ \cos(2\pi \alpha_x x + \theta_x) + \cos(2\pi \alpha_y y + \theta_y) +\cos(2\pi \alpha_z z + \theta_z) \right] b_{x,y,z}^{\dagger} b_{x,y,z},~~\\
H_C=& \sum_{x,y,z=1}^{N_x,N_y,N_z} t a_{x,y,z}^{\dagger} b_{x,y,z} + \text{H.c.},
\end{aligned}
\end{equation}
where $a_{x,y,z}$, $b_{x,y,z}$ ($a_{x,y,z}^{\dagger}$, $b_{x,y,z}^{\dagger}$) are the annihilation (creation) operators of the unit cell $\{x,y,z\}$-th in 3D HN and AA lattices, respectively. $J_{L/R,x/y/z}$ represents the hopping strength of the HN lattice in the $x/y/z$ direction to the left/right, while $t$ is the coupling strength between the two. $N_{x/y/z}$ is the total unit cells in the $x/y/z$-direction. We set $J_{R,x}=J_{R,y}=Je^{g}$, $J_{L,x}=J_{L,y}=Je^{g}$ and $J_{R,z}=J_{R,z}=Je^{g}$, so that when $g>0$, the skin states are stacked towards the ends of the $x$, $y$, $z$-directions.

 \begin{figure}[htbp]
	\centering	\includegraphics[width=13cm]{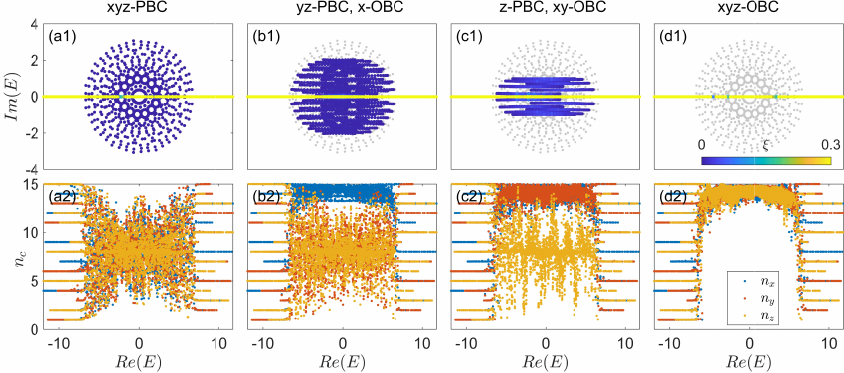}
	\caption{(a1)-(d1) The IPR $\xi$ of all eigenstates in the complex \textcolor{blue}{energy} plane for different boundary conditions, where the gray dots, as a benchmark, represent the eigenvalues under all xyz-direction PBCs. (a2)-(d2) The positions of center-of-mass of all eigenstates under different boundary conditions. The parameters are set as $g=0.5$, $t=0.5$, $N_{x}=N_{y}=N_{z}=15$, $\alpha_{x}=\sqrt{2}$, $\alpha_{y}=(\sqrt{5}-1)/2$, $\alpha_{z}=(\sqrt{13}+3)/2$, and $\lambda=2$. } \label{Fs2}
\end{figure}

Similarly, we gradually open the boundary conditions of different directions and show the changes in the spectrum. Figs.~\ref{Fs2}(a1)-(d1) show the IPR of all eigenvalues in the complex plane for different boundary conditions. Under PBC in the $x$-, $y$- and $z$-directions, similar to 1D and 2D, the same localized Anderson modes will also exist within the point gap formed by the extended states in the 3D case. The gradual opening of the boundaries is also accompanied by a spectral collapse to the real axis, indicating the presence of the non-Hermitian skin effect. Specifically, one can see the variation in the positional center with boundary conditions in Fig.~\ref{Fs2}(a2)-(d2). The results reveal that the localized states within the point gap depend on the boundary conditions. When the boundary conditions are not yet fully open, the hybrid AS dualism mode can emerge. In this case, skin modes are displayed only in certain directions. Under OBCs in all $x$-, $y$-, and $z$-directions, all the localized states within the point gap are transformed into skin modes.


\end{document}